\documentclass[a4paper]{jpconf}
\usepackage{graphicx}
\usepackage{citesort}
\usepackage{color}

\begin{document}
\title{Diagnosing a strong topological insulator by quantum oscillations}

\author{R Ramazashvili, F B\`egue  and P Pujol}
%
\address{
Laboratoire de Physique Th\'eorique  -- IRSAMC, CNRS
and Universit\'e de Toulouse, UPS, F-31062 Toulouse, France }

\ead{revaz.ramazashvili@irsamc.ups-tlse.fr}

\begin{abstract}
We show how quantum oscillation measurements of surface states in
an insulator may allow to diagnose a strong topological insulator and distinguish
it from its weak or topologically trivial counterpart. The criterion is defined by the
parity of the number of fundamental frequencies in the surface-state quantum
oscillation spectrum: an even number of frequencies implies a weak or a
topologically trivial insulator, whereas an odd number points to a strong
topological insulator. We also discuss various aspects and issues related to
applying this criterion in practice.
\end{abstract}





\begin{figure}
\begin{center}\includegraphics[scale=0.5]{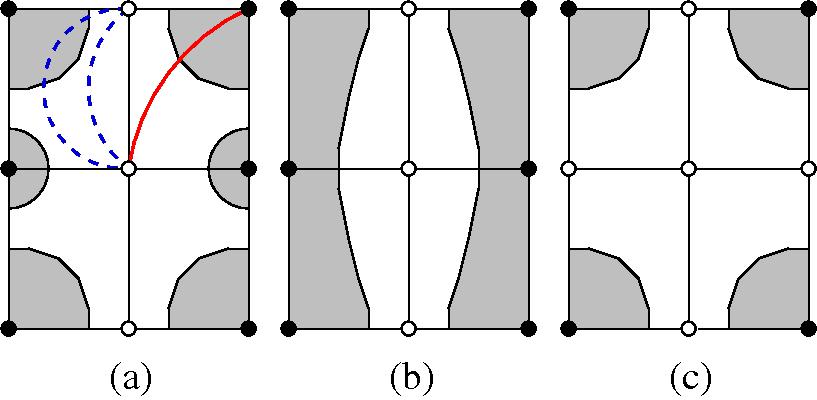}\end{center}
\caption{The first Brillouin zone for the (1 0 1) surface. The small
empty and full circles correspond to the time reversal polarization values $\pi = 1$
and $\pi = -1$, respectively. The panels (a) and (b) show the two generic Fermi
surface configurations, allowed in the same case of a weak topological insulator.
The (a) panel corresponds to two closed Fermi surfaces, and hence two different
fundamental frequencies. The dashed (blue) arcs are examples of paths
connecting two time reversal invariant momenta with the same values of time
reversal polarization. They cross the Fermi surface an even number of times
 -- or not at all. The solid (red) arc is a path connecting two time reversal
invariant momenta with the opposite values of time reversal polarization.
It crosses the Fermi surface an odd number of times.
The (b) panel involves a single open Fermi surface, and thus no quantum oscillations
at all. The case (c) is the generic Fermi surface configuration in the case of a strong
topological insulator, it produces a single fundamental frequency.}
\label{fig:TRP.BZ}
\end{figure}

\section{Introduction}

Study of topological properties of matter has become a frontier of condensed
matter physics. Materials new and old are studied with respect to their
to\-po\-lo\-gy, experimentally and theoretically alike. In particular, due to
great interest in topological insulators, surface states in various insulating
materials are being actively sought and studied. Whenever such states
are experimentally detected, one would like to find out whether they are
of topological or of an accidental origin.

For some materials such as Bi$_2$Te$_3$, Bi$_{2 - \delta}$Ca$_\delta$Te$_3$
and Bi$_{1-x}$Sb$_x$, this question, to a great extent, has been
answered by spin-resolved angle-resolved  photoemission (ARPES) measurements
\cite{hsieh.nature-09,hsieh.science-09}, that were able to detect non-degenerate
helical surface bands with topologically non-trivial spin texture, a key signature of
a topological insulator. On other occasions, the results have been less definitive.
Indeed, in some materials spin-resolved ARPES experiments
turn out to be extremely difficult because of peculiarities of their
surface structure. In other compounds, a significant bulk
conductivity \cite{ando-2013,wright} does not allow transport measurements
to diagnose the topological nature of the material.

An interesting case has recently emerged in the class of the so-called
``Kondo insulators'' \cite{aeppli,coleman}, where a gap in the electron
spectrum opens due to coupling between conduction electrons and
local magnetic moments. After it was pointed out that Kondo insulators may be topologically non-trivial \cite{dzero-2010} while being truly insulating in the bulk
\cite{ando-2013}, experiments on SmB$_6$, a mixed-valence semiconductor
\cite{varma}, have indeed detected surface states, and more theoretical
contributions followed \cite{takimoto,ye-13}. However, even after transport
\cite{wolgast-12,kim-12,Kim,thomas-13}, point contact \cite{zhang-13},
ARPES \cite{xu-13,neupane-13,jiang-13,zhu-13} and scanning tunneling
spectroscopy (STS) \cite{yee-13,rossler-13} experiments,
the available evidence for topological origin of the surface states in SmB$_6$
remained circumstantial -- until very recent spin-resolved
ARPES experiment \cite{SARPESonSMB6} on the ($001$) surface of the
material.

Here, we show how a {\em strong} topological insulator may be identified
via a relatively simple analysis of quantum oscillations due to surface bands.
In SmB$_6$, surface-state quantum oscillations have already been
observed \cite{li-13}. We hope that an analysis of the data \cite{li-13} using the
approach that we propose here may confirm the strong topological insulator nature of this material. We also hope that this approach can be used to diagnose other
potential topological insulators, where spin-resolved ARPES experiments
have not yet been available.

\section{Characterizing a strong topological insulator }

Following the Refs. \cite{fu-2007,kane-08}, consider the time reversal
polarization $\pi=\pm 1$ at the four time reversal invariant momenta of the
surface Brillouin zone. Both a weak topological insulator and a trivial one
will have an even number of time reversal invariant surface momenta with
polarization  $\pi=+1$, while for a strong topological insulator this number
must be odd. A surface state is labeled by its momentum in the surface
Brillouin zone and by a surface band index. Now, in the surface Brillouin
zone, consider a path connecting two time reversal invariant momenta
with the same (different) values of time reversal polarization.
Any such path must contain an even (odd) number of surface states
at the Fermi energy, as shown in the Fig. \ref{fig:TRP.BZ}(a).

Thus we are lead to conclude that both a weak topological insulator and a trivial one
may have either an even non-zero number of different {\em closed} Fermi surfaces
at the sample boundary, as shown in the Fig. \ref{fig:TRP.BZ}(a) -- or none at all.
The latter case also allows for a surface band with an {\em open} Fermi surface,
as shown in the Fig. \ref{fig:TRP.BZ}(b). However, an open Fermi surface does
not contribute to quantum oscillations, a point which is important
for the arguments to follow.

By contrast, as illustrated in the Fig. \ref{fig:TRP.BZ}(c), a topologically protected
surface band in a {\em strong} topological insulator has an odd number of closed Fermi surfaces producing quantum oscillations.

Each closed Fermi surface contributes a fundamental frequency to the spectrum
of quantum oscillations. Thus, barring special cases to be discussed below,
the spectrum of quantum oscillations due to surface states in a {\em strong}
topological insulator contains an odd number of fundamental frequencies.
By contrast, for the quantum oscillation spectrum due to
surface bands in a trivial or a weak topological insulator the number of
fundamental frequencies is even.

In other words, the distinction is not in the number parity of non-degenerate
surface bands: as one can see in the Figs. \ref{fig:TRP.BZ} (b) and (c), this
number may be odd both in a weak and a strong topological insulator. Rather,
it is the odd number of surface bands with a {\em closed} Fermi surface
(that is, an odd number of fundamental frequencies) that distinguishes a
strong topological insulator from its weak or topologically trivial counterpart.

The arguments above assume that the physical picture of a non-interacting
topological insulator, developed in the Refs. \cite{fu-2007,kane-08}, holds
for the materials in question, in spite of the presence of electron correlations.

In practice, one should first determine the bulk or surface nature of the studied
pocket of carriers by examining the dependence of the quantum oscillations on 
the direction of the applied magnetic field \cite{shoenberg}. Next, one shall find 
out whether the observed surface band is degenerate or not. This amounts to 
checking whether it produces one or two fundamental frequencies. If two 
fundamental frequencies $F_1$ and $F_2$ were to be resolved, they would be 
seen as two separate peaks in the Fourier transform of the quantum oscillation 
data. However, if the $F_1$ and $F_2$ were too close to be resolved as separate 
peaks, they could still be detected via the emerging modulation factor (beats) 
\cite{shoenberg}: 
\begin{equation}
\mathcal{R} =
\cos
\left[
\pi \frac{\delta F}{H}
\right],
\label{eq:modulation}
\end{equation}
where $\delta F = F_1 - F_2$.
For this, there are two distinct possibilities.

The first one corresponds to the Zeeman splitting,
where the Eq. (\ref{eq:modulation}) reduces to
\begin{equation}
\mathcal{R} =
\cos
\left[
\pi \frac{\delta \mathcal{E}_Z}{\Omega_0}
\right].
\end{equation}
Here, $\delta \mathcal{E}_Z$ is the Zeeman splitting, $\Omega_0$
the cyclotron frequency, and the $\mathcal{R}$ is usually referred to
as the spin reduction factor. Since both the $\delta \mathcal{E}_Z$ and
the $\Omega_0$ are proportional to the field strength, the $\mathcal{R}$
depends only on the field orientation with respect to the surface, and
vanishes for particular orientations known as "spin zeros" \cite{shoenberg}.

The second possibility corresponds to an intrinsic spin-orbit coupling, where
the splitting $\delta F$ in the Eq. (\ref{eq:modulation}) does not vanish in the
$H \rightarrow 0$ limit. By contrast with the case of Zeeman splitting, here the
amplitude modulation is a function of both the field strength and its orientation.

One may argue that in certain special cases two non-degenerate surface bands 
may give rise to a single fundamental frequency, and thus a trivial insulator 
would look like a topological one. Indeed, this could happen in a system with an 
anomalously weak Zeeman effect, that would not allow for the separation of spin 
degenerate surface states, even in the presence of a magnetic field. The same
problem may arise if two non-coincident Fermi surfaces can be mapped on
each other by a symmetry transformation: in this case, the two fundamental
frequencies coincide. In the former case, a simple estimation of the Zeeman
splitting and of the cyclotron frequency would allow one to determine the intensity
of the magnetic field needed to resolve two distinct frequencies.
By contrast, the latter case is more delicate: here, the degeneracy 
of the two fundamental frequencies is protected by symmetry.

\section{An example: the case of SmB$6$}

An interesting candidate topological material, SmB6, appeared recently in the class 
of the so-called ``Kondo Insulators". A beautiful transport experiment \cite{wolgast-12} 
successfully distinguished bulk-dominated from surface-dominated conduction: 
it was found that, as the material is cooled below 4 K, it exhibits a crossover 
from bulk to surface conduction. Soon, a quantum oscillation experiment 
\cite{li-13} probed the electron states, bound to the (101) surface. The reported 
torque data exhibit two fundamental frequencies, attributed to the two pockets, 
named $\alpha$ and $\beta$. The two frequencies are shown in the Fig. 2 (b) 
of the Ref. \cite{li-13}: only the $\beta$-frequency depends on the magnetic field 
orientation in a way consistent with a (101) surface state behavior 
(see the Fig. 3 (a) of the Ref. \cite{li-13}).

On the other hand, the $\alpha$-pocket has been argued to arise from the less clean and polar ($001$) surface\cite{li-13,Frantzeskakis,zhu-13}. Another explanation would be that this pocket is of a bulk origin, as  the $\alpha$-frequency depends on the field orientation very weakly if at all, which is hardly consistent with surface character of the carrier states, but rather suggestive of the said pocket being three-dimensional. As we wish to concentrate on the ($101$) surface, we disregard
the $\alpha$-pocket as extrinsic to the physics at hand.

Our task is thus to find out whether the $\beta$ oscillations
represent one or two fundamental frequencies.
Unfortunately, the Ref. \cite{li-13} presents no explicit analysis
of amplitude modulations or spin zeros as a function of field orientation and
magnitude, and thus we do not know whether these are actually present in
the data. However, the analysis we propose offers a direct and robust test
of whether a material (in the present case, SmB$_6$) is a strong topological
insulator. We argue here, that, to be consistent with spin-resolved ARPES measurement,
an analysis of the data should point out that the $\beta$ oscillations represent a unique fundamental frequency.

\begin{figure}
\begin{center}\includegraphics[scale=0.4]{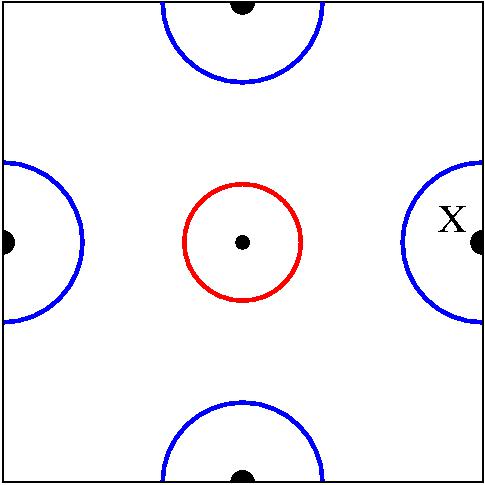}\end{center}
\caption{The first Brillouin zone for the (1 0 0) surface of SmB$_6$. As shown
by  \cite{ye-13}, the surface pockets centered at the point X and its $\pi/2$ rotation
counterpart yield degenerate fundamental frequencies. The pocket centered
at the point $\Gamma$ of the surface Brillouin zone yields a different fundamental
frequency. By the fundamental frequency count, the model would thus mimic
as a week topological insulator while in fact corresponding to a strong one.
}
\label{fig:degen}
\end{figure}

It is important to note that if a quantum oscillation experiment were done on the
($001$) surface, the results would probably be less definitive than on the ($101$) 
surface. Indeed, it has been shown by ARPES experiments \cite{neupane-13,xu-13,jiang-13,SARPESonSMB6},
that on this surface, there are three closed Fermi surfaces, shown 
in the Fig. \ref{fig:degen}. 
According to \cite{ye-13}, the two Fermi pockets centered at the point X of the surface 
Brillouin zone and its image upon rotation by $\pi/2$ around $\Gamma$ produce 
identical fundamental frequencies. The counting of fundamental frequencies (two) 
would thus point to a weak topological insulator, while the number of closed Fermi 
surfaces is odd and, in fact, the model describes a strong topological insulator. 
This spurious degeneracy can be removed by subjecting the sample to an uniaxial 
pressure. However, this reduces the practical simplicity of the criterion we propose.

\section{Conclusion}

To summarize, we proposed a method of diagnosing a strong topological
insulator by counting the number of fundamental frequencies, observed
in magnetic quantum oscillations due to surface states. We expect the
method to work well except for cases of surface-state Fermi surfaces
that are degenerate by symmetry. How does the method we proposed
compare with its counterparts?

Recently, quantum oscillation data have been used to verify the Dirac dispersion
of carriers via analysis of the Berry phase $\gamma$ of the oscillations as in
$\cos \left[ 2 \pi \frac{F_1}{H} + \pi + \gamma \right]$ -- both in graphene \cite{zhang}
and in topological insulators \cite{xiong}. However, such an analysis comes with
its own challenges, described in the Section 8.3 of the Ref. \cite{ando-2013},
and in the Ref.  \cite{wright}. By contrast, counting the fundamental frequencies
of surface-state quantum oscillations, as we propose, appears to present a much
simpler task.

The Berry phase analysis and the counting of fundamental frequencies may be
viewed as complementary to each other, in the following sense. The $\gamma = \pi$
does not, by itself, prove the topological nature of the surface states, to which
the Dirac spectrum in single-layer graphene is an example. Another example
is the Rashba Hamiltonian
$\mathcal{H}_R = \alpha ({\bf n} \cdot {\bf p} \times {\bf \sigma}) + {\bf p}^2 / 2m$
of surface states in an inversion layer \cite{bychkov}: it is not related to non-trivial
band topology in the bulk, yet in the vicinity of zero momentum it is equivalent to
the Hamiltonian of a surface Dirac branch in a topological insulator.
However, once a material has been independently shown to be a topological insulator,
finding $\gamma = \pi$ serves as a confirmation of the Dirac character of the surface
state spectrum.

Note that, compared with ARPES, the present approach
allows to routinely resolve the Zeeman-split quantum oscillation frequencies,
whereas even for the state-of-the-art ARPES with its current energy resolution
of several meV \cite{hsieh.nature-09,hsieh.science-09,damascelli} this remains
a challenge.

To summarize, we pointed out how quantum oscillation experiments
may allow to distinguish surface states in a strong topological
insulator from those in its weak or topologically trivial counterpart. As an
illustration, we discussed recent experiments  \cite{li-13} on SmB$_6$.


\section*{References}

\bibliography{biblio}


\end{document}